\documentclass{Interspeech}

\interspeechcameraready

\title{A Chinese Heart Failure Status Speech Database with Universal and Personalised Classification}

\author[affiliation={1}]{Yue}{Pan}
\author[affiliation={2}]{Liwei}{Liu}
\author[affiliation={2}]{Changxin}{Li}
\author[affiliation={3}]{Xingyao}{Wang}
\author[affiliation={1}]{Yili}{Xia}
\author[affiliation={4}]{Hanyue}{Zhang}
\author[affiliation={4}]{Ming}{Chu}

\affiliation{School of Information Science and Technology}{Southeast University}{China}
\affiliation{Advanced Computing and Storage Laboratory, 2012 Laboratories}{Huawei Technologies Co. Ltd.}{}
\affiliation{Institute of High Performance Computing}{A*STAR}{Singapore}
\affiliation{Taizhou School of Clinial Medicine}{Nanjing Medical University}{China}
\email{230228208@seu.edu.cn, liuliwei5@huawei.com, chuming@njmu.edu.cn}

\usepackage{comment}
\usepackage{multirow}
\usepackage{cases}

\begin{document}

\maketitle

\begin{abstract}
    
    Speech is a cost-effective and non-intrusive data source for identifying acute and chronic heart failure (HF). However, there is a lack of research on whether Chinese syllables contain HF-related information, as observed in other well-studied languages. This study presents the first Chinese speech database of HF patients, featuring paired recordings taken before and after hospitalisation. The findings confirm the effectiveness of the Chinese language in HF detection using both standard 'patient-wise' and personalised 'pair-wise' classification approaches, with the latter serving as an ideal speaker-decoupled baseline for future research. Statistical tests and classification results highlight individual differences as key contributors to inaccuracy. Additionally, an adaptive frequency filter (AFF) is proposed for frequency importance analysis. The data and demonstrations are published at \url{https://github.com/panyue1998/Voice_HF}.
\end{abstract}

\section{Introduction}
Heart failure (HF) is a progressive condition characterised by a decline in the heart's ability to pump blood effectively, affecting 26 million individuals worldwide \cite{amir2022remote}. Speech analysis offers a cost-effective and non-intrusive method for detecting HF-related laryngeal edema in its early stages \cite{murton2023acoustic}, providing an alternative to conventional diagnostic techniques, such as X-ray, echocardiography, and angiography. Previous research has explored various speech tasks for HF detection, including vowels and word articulation \cite{murton2017acoustic}, sentence reading \cite{amir2022remote, murton2017acoustic, firmino2023heart}, and short paragraph recitations \cite{murton2017acoustic, reddy2021automatic, priyasad2022detecting}. These studies have been conducted in multiple languages, such as English \cite{murton2023acoustic, murton2017acoustic}, Finnish \cite{reddy2021automatic, mittapalle2022glottal}, and Portuguese \cite{firmino2023heart}, while some supported a mix of different languages \cite{amir2022remote}. 

Since HF-related changes in speech features are subtle and can be overshadowed by individual variations, a paired database that captures multiple conditions of the same patient is essential for extracting pathological information. One of the earliest studies in this direction, conducted by Murton et al. \cite{murton2017acoustic}, analysed speech from ten hospitalised HF patients, revealing detectable trends in phonation and respiration parameters between admission (wet) and discharge (dry) conditions. Similarly, Amir et al. \cite{amir2022remote} observed vocal alterations between wet and dry states using a mobile application designed for speech analysis. Both studies highlighted intra-patient variability as a potential confounding factor. A subsequent study expanded on \cite{murton2017acoustic} by incorporating a larger patient cohort and additional biomarkers, demonstrating a positive correlation between speech features and discharge probability through logistic regression analysis \cite{murton2023acoustic}. However, these studies did not systematically assess the impact of individual differences on classification accuracy, nor did they establish baselines to evaluate such effects. 

To address these gaps, this study introduces a pair-wise classification approach alongside the standard patient-wise classification method, leveraging a newly constructed, large-scale Chinese paired speech database. A summary of related studies is shown in Table \ref{previous}. Our dataset is among the most extensive, particularly in terms of paired speech samples collected before and after medical intervention. While no public dataset is currently available, we intend to release high-level feature data while preserving patient privacy. Previous studies primarily focused on Indo-European and Afroasiatic languages, largely neglecting Sino-Tibetan languages. Although HF detection is generally considered content-independent, linguistic characteristics---such as those specific to Chinese \cite{liu2009formant, li2018parkinson}---may influence pathological speech markers. Consequently, features identified in one language may not necessarily apply to another. 

Furthermore, prior studies reported varied classification accuracies without thoroughly analysing the sources of model errors. This work hypothesises that individual differences significantly contribute to classification inaccuracies, a claim supported by statistical tests and our proposed pair-wise classification framework, which also serves as a robust baseline for future research. Additionally, we introduce an adaptive frequency filter (AFF) for frequency importance analysis in time-frequency sequential models. The key contributions of this work include:
\begin{itemize}
    \item Development of the first Chinese speech database for HF detection with paired samples, achieving high classification performance. 
    \item Introduction of a 'pair-wise' classification approach as a speaker-independent baseline, identifying individual variations as a primary source of classification inaccuracy.  
    \item Design of an AFF for frequency importance analysis. 
\end{itemize}

\begin{table*}[]
\centering
\begin{tabular}{lllllll}
Study                         & Year & Language                                                              & Patients & \begin{tabular}[c]{@{}l@{}}Data \\ Collection\end{tabular} & \begin{tabular}[c]{@{}l@{}}Data\\ Made Public\end{tabular}                                    & \begin{tabular}[c]{@{}l@{}}Best Result\\ (Accuracy)\end{tabular} \\ \hline
\cite{murton2017acoustic}    & 2017 & English                                                               & 10       & Paired                                                     & No                                                                                       & -                                                                \\
\cite{reddy2021automatic}    & 2021 & Finnish                                                               & 45       & Single                                                     & No                                                                                       & 81.5                                                             \\
\cite{mittapalle2022glottal} & 2022 & Finnish                                                               & 45       & Single                                                     & No                                                                                       & -                                                                \\
\cite{amir2022remote}        & 2022 & \begin{tabular}[c]{@{}l@{}}Hebrew,  Arabic, Russian\end{tabular} & 40       & Paired                                                     & No                                                                                       & -                                                                \\
\cite{priyasad2022detecting}        & 2022 & \begin{tabular}[c]{@{}l@{}}English\end{tabular} & 74       & Single                                                     & No                                                                                       & 93.7                                                                \\
\cite{firmino2023heart}      & 2023 & Portuguese                                                            & 142      & Single                                                     & No                                                                                       & 91.9                                                             \\
\cite{murton2023acoustic}    & 2023 & English                                                               & 52       & Multiple                                                   & No                                                                                       & 69.0                                                             \\
Ours                          &      & Chinese                                                               & 127      & Paired                                                     & \begin{tabular}[c]{@{}l@{}}High-level features/\\Full data by request\end{tabular} & See results                                                      \\ \hline
\end{tabular}
\caption{Summary of similar studies}\label{previous}
\end{table*}

\begin{figure*}[h]%
\includegraphics[width=1\textwidth,trim=0 180 0 0,clip]{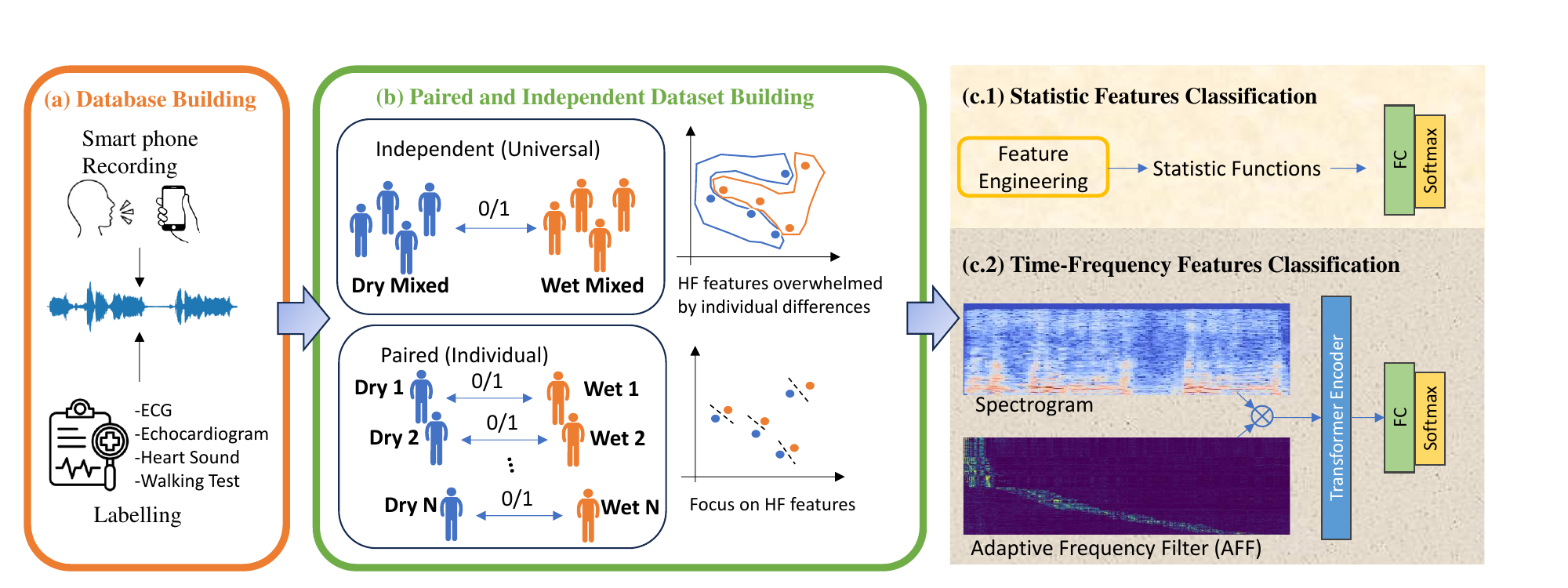}
\caption{Overall structure of the project.}\label{structure}
\end{figure*}

In Section \ref{Method}, the proposed methods and experimental setup are detailed. The results and discussion are presented  in Section \ref{Results}, while the conclusion and suggestions for future work are provided in Section \ref{Conclusion}. A summary of related studies is illustrated in Figure \ref{structure}.   

\section{Methods}
\label{Method}
\subsection{Data acquisition}
\label{Data Acquisition}
This study involves a total of 127 patients from our partner hospital, admitted for acute HF treatments. The recordings were collected using a standard smartphone, handheld by the staff, at a sampling rate of 22,050 Hz. Each patient participated in two data collection sessions, one before and after hospitalisation. The male-to-female ratio is 0.61:0.39, with an average age of 68 and a standard deviation of 13. 

Medical professionals assessed the patients' conditions based on the New York Heart Association (NYHA) functional classification for HF before and after hospitalisation, considering clinical tests, such as electrocardiogram, echocardiography, and walking tests \cite{giannitsi20196}. Patients who exhibited improvement in NYHA levels were considered relevant for the study. 

Participants were required to complete four speech tasks, categorised into short and long sentences. In Chinese, consonants exhibit the discrimination between voiced and unvoiced sounds. These are further classified into fully voiced (r), partially voiced (m, n, l, j, w), fully unvoiced (b, d, g, s, x, z), and partially unvoiced (p, t, k, c, q) \cite{li2018parkinson}. Three short sentences were selected to capture both voiced and unvoiced sounds in Chinese. For the long sentence task, participants were asked to count from 1 to 60 in Chinese, ensuring the inclusion of both voiced and unvoiced sounds. Vowels (a, i, u) were present in all four tasks. Details of these tasks are presented in Table \ref{tasks}. 

Every recording was manually labelled according to its respective task. Unrelated, incomplete, and erroneous samples, as well as those from patients with other underlying conditions affecting speech, were excluded, leaving a total of 117 patients. The final number of recordings for each task after data cleansing is shown in Table \ref{train_test_amount}.

\begin{table*}[]
\centering
\begin{tabular}{llllll}
task             & abbr. & \begin{tabular}[c]{@{}l@{}}Chinese Pingyin\\ (phonetic symbols)\end{tabular} & Consonants    & \begin{tabular}[c]{@{}l@{}}Consonants\\ Type\end{tabular} & \begin{tabular}[c]{@{}l@{}}Average Length \\ of Recordings\end{tabular} \\ \hline
Short Sentence 1 & pg    & shan dong de ping guo you da you tian                                        & s, d, p, g, t & unvoiced                                                  & 5.1s                                                                    \\
Short Sentence 2 & mm    & ni you yi ge mei li de mei mei                                               & m, n, l       & voiced                                                    & 4.6s                                                                    \\
Short Sentence 3 & mlh   & hao yi duo mei li de mo li hua                                               & m, l          & voiced                                                    & 4.7s                                                                    \\
Long Sentence    & c     & (numbers 1-60)                                                               & r, l, j, b, q & both                                                      & 27.2s                                                                   \\ \hline
\end{tabular}
\caption{Speach tasks conducted.}\label{tasks}
\end{table*}

\begin{table}[h]
\centering
\begin{tabular}{llllll}
                                                                                  & Tasks  & \textbf{pg} & \textbf{mm} & \textbf{mlh} & \textbf{c} \\ \hline
\multirow{3}{*}{\begin{tabular}[c]{@{}l@{}}Training\\ (Individuals)\end{tabular}} & male   & 46          & 44          & 38           & 47         \\
                                                                                  & female & 28          & 21          & 23           & 30         \\
                                                                                  & total  & 74          & 65          & 61           & 77         \\ \hline
\multirow{3}{*}{\begin{tabular}[c]{@{}l@{}}Testing\\ (Individuals)\end{tabular}}  & male   & 22          & 17          & 19           & 20         \\
                                                                                  & female & 11          & 5           & 6            & 11         \\
                                                                                  & total  & 33          & 22          & 25           & 31         \\ \hline
\end{tabular}
\caption{Number of individuals used for training and testing in each task. Task name abbreviations are provided in Table \ref{tasks}.}\label{train_test_amount}
\end{table}

\subsection{Feature extraction and selection}
\label{Feature Extraction and Selection}
This study performed feature extraction using openSMILE \cite{eyben2010opensmile}, version 2.5.0. Two feature sets were utilised: GeMAPS \cite{eyben2015geneva} and ComParE 2016 \cite{schuller2016interspeech}. To identify the most relevant features associated with hospitalisation conditions, paired and independent t-tests were conducted between admission and discharge groups. Table \ref{feature selection} presents the number of selected features where $p\leq 0.05$ for each task. This process was repeated for each task and both feature sets, considering female, male, and combined groups separately. The selected feature data were then used to train a three-layer fully connected neural network for HF detection, as illustrated in part c.1 in Figure \ref{structure}.

\begin{table}[h]
\begin{tabular}{lllrrrr}
\multirow{2}{*}{\begin{tabular}[c]{@{}l@{}}Feature \\ Set\end{tabular}} & \multirow{2}{*}{Sex}    & \multirow{2}{*}{T-test} & \multicolumn{4}{l}{Tasks}                                                                                                             \\
                               &                         &                         & \multicolumn{1}{l}{\textbf{pg}} & \multicolumn{1}{l}{\textbf{mm}} & \multicolumn{1}{l}{\textbf{mlh}} & \multicolumn{1}{l}{\textbf{c}} \\ \hline
\multirow{6}{*}{A}    & \multirow{2}{*}{male}   & ind                     & 2                               & 7                               & 4                                & 3                              \\
                               &                         & pair                    & 8                               & 12                              & 5                                & 6                              \\ \cline{2-7} 
                               & \multirow{2}{*}{female} & ind                     & 5                               & 7                               & 2                                & 3                              \\
                               &                         & pair                    & 15                              & 9                               & 11                               & 12                             \\ \cline{2-7} 
                               & \multirow{2}{*}{all}    & ind                     & 11                              & 10                              & 5                                & 7                              \\
                               &                         & pair                    & 23                              & 20                              & 16                               & 15                             \\ \hline
\multirow{6}{*}{B} & \multirow{2}{*}{male}   & ind                     & 248                             & 241                             & 187                              & 467                            \\
                               &                         & pair                    & 392                             & 378                             & 352                              & 914                            \\ \cline{2-7} 
                               & \multirow{2}{*}{female} & ind                     & 332                             & 327                             & 311                              & 467                            \\
                               &                         & pair                    & 562                             & 461                             & 440                              & 996                            \\ \cline{2-7} 
                               & \multirow{2}{*}{all}    & ind                     & 326                             & 287                             & 212                              & 901                            \\
                               &                         & pair                    & 618                             & 490                             & 434                              & 1483                           \\ \hline
\end{tabular}
\caption{Number of features selected by paired (pair) and independent (ind) t-tests, where $p\leq 0.05$. A: ComParE\_2016, B: eGeMAPSv02. Task name abbreviations are provided in Table \ref{tasks}.}\label{feature selection}
\end{table}

\subsection{Pair-wise classification}
\label{Classification Methods}
In this work, we propose the 'pair-wise' classification scheme as a baseline for the ideal speaker-decoupled case. In the pair-wise scenario, we aim to simultaneously feed both the wet and dry data points of a single patient into the classifier, which should then determine which is wet and which is dry. In other words, there is an additional by-patient normalisation compared with the raw data. The commonality with the standard scheme is that the training and testing sets are divided according to patient ID, ensuring that the data points in the test set are from patients previously unseen by the classifier, so they are both speaker-independent. 

Consider the selected feature vectors of a given patient as $A^i=[a_1^i, a_2^i, ... ,a_n^i]$ (wet) and $B^i=[b_1^i, b_2^i, ... ,b_n^i]$ (dry), where \textit{i} is the patient ID and \textit{n} is the feature ID. For the pair-wise scheme, we first randomly generate a 0/1 label. If the label is 1, a combined vector is formed where the 'wet' vector is subtracted from the 'dry' one, and vice versa:
\begin{subnumcases} {\label{weqn} X_{train/test}^i=}
A^i - B^i & for $label=1$ \\
B^i - A^i & for $label=0$ 
\end{subnumcases}

In this way, the result is a comparison within a single patient's data, making it immune to the domain difference caused by inherent individual variations in speech voice. While this scheme may not be as practical as the standard patient-wise scheme in real-world applications---since it would require a known normal state for a given individual--- it serves as a useful baseline reference in cases where individual differences are decoupled from pathological features.

The standard patient-wise scheme follows a typical classification approach, where the training and test sets contain a mixture of 'wet' and 'dry' vectors, each treated as a standalone data point. The standard scheme was conducted separately for female, male, and combined groups.

\subsection{Adaptive frequency filter (AFF)}
The AFF (part c.2 in Figure \ref{structure}) is primarily designed for the visualisation of frequency importance. This technique is inspired by \cite{yang2023attention}. While \cite{yang2023attention} operates in the time domain, our AFF is directly applied in the frequency domain. The AFF is a trainable linear projection matrix applied to the time-frequency data before sequential encoding, and in this case, transformer encoders. It has a dimension of $(d_{freq}, d_{new})$, where $d_{freq}$ represents the input dimension of the frequency axis, and $d_{new}$ is a user-defined target dimension. It is applied to the spectrogram (S) as:
\begin{equation}
    \mathbf{S}\left( {T,d_{freq}} \right) \times \mathbf{A}\mathbf{F}\mathbf{F}~\left( {d_{freq},d_{new}} \right) = \mathbf{F}\left( T,d_{new} \right)
\end{equation}

To obtain the filtered feature map (F), the AFF will have $d_{new}$ filters, with each filter having trainable attention across all $d_{freq}$ dimensions. The trainable AFF is initialised as an MFCC filter bank. To ensure that each filter focuses on a relevant frequency band and converges towards a Butterworth-style filter bank, we trim the high attention values outside a specific frequency range around the current highest position after every few epochs. 

\section{Results and discussion}
\label{Results}
Table \ref{result} presents the patient-wise and pair-wise classification results, as outlined in part c.1 of Figure \ref{structure}. We report the F1 score, which is calculated as follows: 
\begin{equation}
    F1 = \frac{2 \times precision \times recall}{precision + recall}
\end{equation}

\begin{table}[b!]
\renewcommand\arraystretch{0.70}
\begin{tabular}{lllllll}
\multirow{2}{*}{Task}         & \multirow{2}{*}{\begin{tabular}[c]{@{}l@{}}Feature \\ Set\end{tabular}} & \multirow{2}{*}{\begin{tabular}[c]{@{}l@{}}Feature \\ Selection\end{tabular}} & \multicolumn{4}{l}{F1 (\%)}                                                                                                             \\
                              &                                                                         &                                                                               & Female                         & Male                  & All                   & \begin{tabular}[c]{@{}l@{}}Pair-\\ wise\end{tabular} \\ \hline
\multirow{8}{*}{\textbf{pg}}  & \multirow{4}{*}{A}                                                      & \multirow{2}{*}{ind}                                                          & \multirow{2}{*}{68.1}          & \multirow{2}{*}{54.2} & \multirow{2}{*}{54.5} & \multirow{2}{*}{\textbf{68.6}}                       \\
                              &                                                                         &                                                                               &                                &                       &                       &                                                      \\
                              &                                                                         & \multirow{2}{*}{pair}                                                         & \multirow{2}{*}{59.0}          & \multirow{2}{*}{59.0} & \multirow{2}{*}{56.0} & \multirow{2}{*}{\textbf{61.6}}                       \\
                              &                                                                         &                                                                               &                                &                       &                       &                                                      \\ \cline{2-7} 
                              & \multirow{4}{*}{B}                                                      & \multirow{2}{*}{ind}                                                          & \multirow{2}{*}{\textbf{90.9}} & \multirow{2}{*}{86.4} & \multirow{2}{*}{72.7} & \multirow{2}{*}{89.6}                                \\
                              &                                                                         &                                                                               &                                &                       &                       &                                                      \\
                              &                                                                         & \multirow{2}{*}{pair}                                                         & \multirow{2}{*}{\textbf{90.8}} & \multirow{2}{*}{79.5} & \multirow{2}{*}{68.1} & \multirow{2}{*}{88.6}                                \\
                              &                                                                         &                                                                               &                                &                       &                       &                                                      \\ \hline
\multirow{8}{*}{\textbf{mm}}  & \multirow{4}{*}{A}                                                      & \multirow{2}{*}{ind}                                                          & \multirow{2}{*}{58.3}          & \multirow{2}{*}{58.7} & \multirow{2}{*}{56.8} & \multirow{2}{*}{\textbf{68.8}}                       \\
                              &                                                                         &                                                                               &                                &                       &                       &                                                      \\
                              &                                                                         & \multirow{2}{*}{pair}                                                         & \multirow{2}{*}{49.5}          & \multirow{2}{*}{55.8} & \multirow{2}{*}{47.7} & \multirow{2}{*}{\textbf{76.3}}                       \\
                              &                                                                         &                                                                               &                                &                       &                       &                                                      \\ \cline{2-7} 
                              & \multirow{4}{*}{B}                                                      & \multirow{2}{*}{ind}                                                          & \multirow{2}{*}{89.9}          & \multirow{2}{*}{85.2} & \multirow{2}{*}{68.1} & \multirow{2}{*}{\textbf{96.4}}                       \\
                              &                                                                         &                                                                               &                                &                       &                       &                                                      \\
                              &                                                                         & \multirow{2}{*}{pair}                                                         & \multirow{2}{*}{89.9}          & \multirow{2}{*}{67.6} & \multirow{2}{*}{61.2} & \multirow{2}{*}{\textbf{96.4}}                       \\
                              &                                                                         &                                                                               &                                &                       &                       &                                                      \\ \hline
\multirow{8}{*}{\textbf{mlh}} & \multirow{4}{*}{A}                                                      & \multirow{2}{*}{ind}                                                          & \multirow{2}{*}{48.5}          & \multirow{2}{*}{57.8} & \multirow{2}{*}{61.3} & \multirow{2}{*}{\textbf{71.3}}                       \\
                              &                                                                         &                                                                               &                                &                       &                       &                                                      \\
                              &                                                                         & \multirow{2}{*}{pair}                                                         & \multirow{2}{*}{41.2}          & \multirow{2}{*}{59.8} & \multirow{2}{*}{53.8} & \multirow{2}{*}{\textbf{73.0}}                       \\
                              &                                                                         &                                                                               &                                &                       &                       &                                                      \\ \cline{2-7} 
                              & \multirow{4}{*}{B}                                                      & \multirow{2}{*}{ind}                                                          & \multirow{2}{*}{91.6}          & \multirow{2}{*}{78.9} & \multirow{2}{*}{76.0} & \multirow{2}{*}{\textbf{91.7}}                       \\
                              &                                                                         &                                                                               &                                &                       &                       &                                                      \\
                              &                                                                         & \multirow{2}{*}{pair}                                                         & \multirow{2}{*}{74.8}          & \multirow{2}{*}{76.3} & \multirow{2}{*}{71.8} & \multirow{2}{*}{\textbf{91.2}}                       \\
                              &                                                                         &                                                                               &                                &                       &                       &                                                      \\ \hline
\multirow{8}{*}{\textbf{c}}   & \multirow{4}{*}{A}                                                      & \multirow{2}{*}{ind}                                                          & \multirow{2}{*}{54.2}          & \multirow{2}{*}{60.0} & \multirow{2}{*}{62.7} & \multirow{2}{*}{\textbf{67.3}}                       \\
                              &                                                                         &                                                                               &                                &                       &                       &                                                      \\
                              &                                                                         & \multirow{2}{*}{pair}                                                         & \multirow{2}{*}{54.5}          & \multirow{2}{*}{55.0} & \multirow{2}{*}{64.5} & \multirow{2}{*}{\textbf{61.0}}                       \\
                              &                                                                         &                                                                               &                                &                       &                       &                                                      \\ \cline{2-7} 
                              & \multirow{4}{*}{B}                                                      & \multirow{2}{*}{ind}                                                          & \multirow{2}{*}{68.1}          & \multirow{2}{*}{80.0} & \multirow{2}{*}{74.2} & \multirow{2}{*}{\textbf{95.4}}                       \\
                              &                                                                         &                                                                               &                                &                       &                       &                                                      \\
                              &                                                                         & \multirow{2}{*}{pair}                                                         & \multirow{2}{*}{72.7}          & \multirow{2}{*}{77.5} & \multirow{2}{*}{74.1} & \multirow{2}{*}{\textbf{96.5}}                       \\
                              &                                                                         &                                                                               &                                &                       &                       &                                                      \\ \hline
\multicolumn{3}{l}{\textbf{Average}}                                                                                                                                                    & 68.9                           & 68.2                  & 64.0                  & \textbf{80.9}                                        \\ \hline
\end{tabular}
\caption{Classification results of fully connected classifiers. A: ComParE\_2016, B: eGeMAPSv02. Task name abbreviations are provided in Table \ref{tasks}.}\label{result}
\end{table}

\begin{figure*}[t]%
\centering
\includegraphics[width=1.0\textwidth,trim=0 97 90 0,clip]{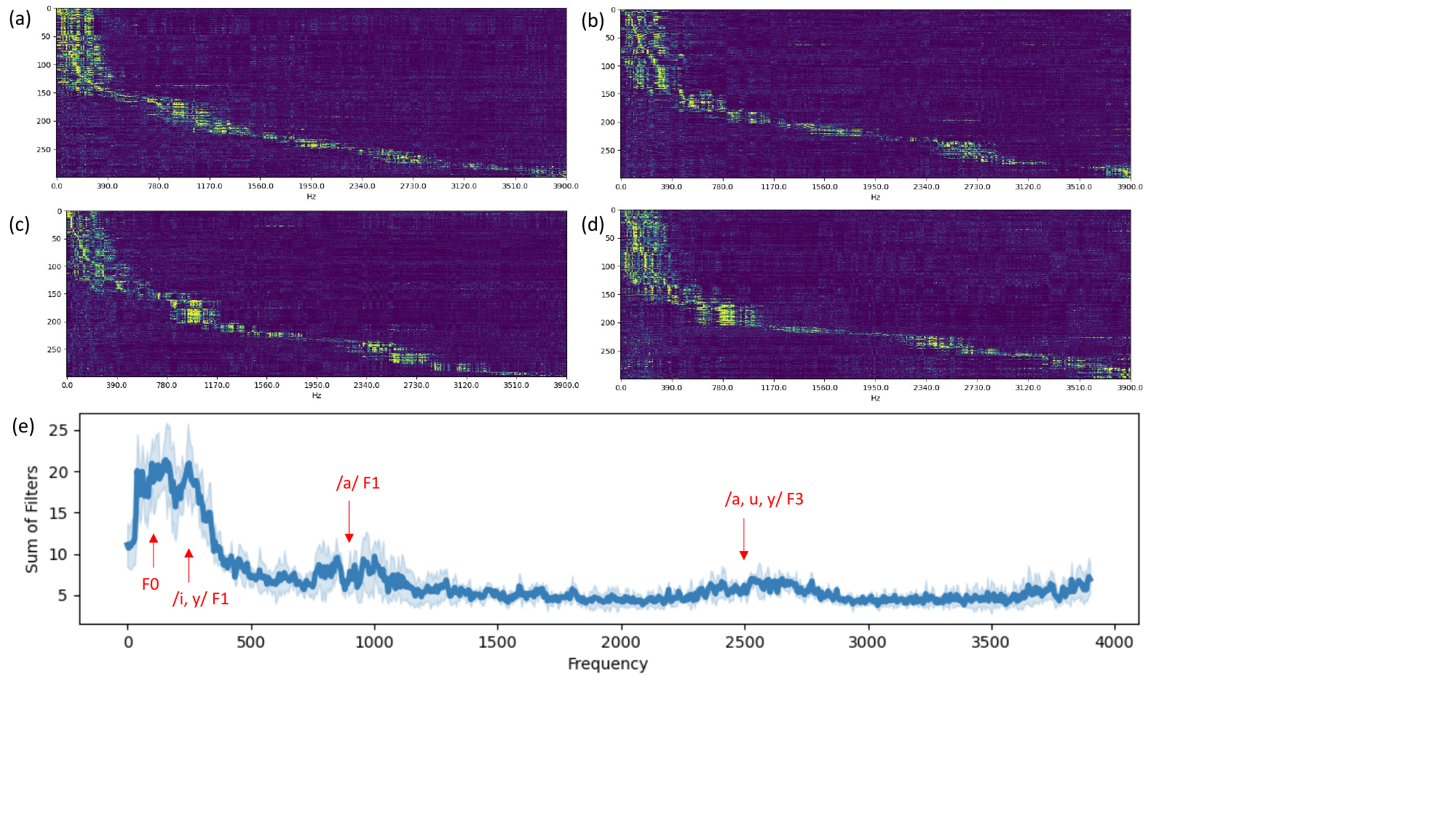}
\caption{AFF frequency analysis for the tasks: pg(a), mm(b), mlh(c), c(d); and the sum of frequency dimensions, average, and standard deviation across the four tasks (e). The positions of the fundamental frequency (F0) and the resonant frequencies of vowels are marked in (e). For task name abbreviations, refer to Table \ref{tasks}.}\label{AFF_result}
\end{figure*}

Overall, the ComParE\_2016 feature set outperformed eGeMAPSv02, though it comes with a considerably larger feature size. The highest performance was observed in the pair-wise scheme of the 'mm' task using the ComParE\_2016 feature set, achieving an F1 score of 0.964. The pair-wise scheme generally outperformed the patient-wise scheme across nearly all settings, including the average score. This suggests that personal differences substantially influence model accuracy, as expected, with the pair-wise scheme remaining unaffected by inter-patient variations due to its focus on intra-patient comparisons. This assumption is further supported by statistical tests. From Table \ref{feature selection}, paired t-tests consistently identified more significant features than independent t-tests, indicating that while there are changes before and after hospitalisation, these differences are smaller than the inherent variations between patients. As a result, the overall difference between HF and normal groups becomes less pronounced.

Although previous studies commonly train separate models for male and female groups, our results show that this approach slightly improves performance (about +4\% on average compared with the combined group). However, it cannot completely eliminate the influence of personal differences, as evidenced by the remaining 12\% difference compared to the pair-wise baseline.  

The frequency analysis of the AFF (part c.2 in Figure \ref{structure}) is illustrated in Figure \ref{AFF_result}. Several significant frequency areas are identified, with the low-frequency range below 250 Hz corresponding to the fundamental frequency (F0). The AFF also captured several vowel formants, such as /a/ (F1 $\sim800Hz$), /i, y/ (F1 $\sim300Hz$) and /u/ \cite{liu2009formant}. Notably, in Figure \ref{AFF_result}(b), the 800 Hz area is less pronounced for the 'mm' task, likely due to the absence of the vowel /a/.

\section{Conclusion}
\label{Conclusion}
This study presents the first large-scale Chinese paired speech dataset for HF detection. A 'pair-wise' classification scheme, decoupling personal differences, was proposed as an ideal baseline. Two feature sets were employed for feature extraction, and fully connected models were used for classification. The classification results highlight personal differences as a major factor affecting model accuracy, a challenge that cannot be fully addressed by the common practice of training separate models for male and female groups. The frequency analysis using the AFF identified several vowel formants as significant in HF detection.

Future work could improve upon this study in several ways. First, although the 'pair-wise' scheme serves as an ideal reference for speaker-irrelevant cases, it may not be practical in real-world applications, as it requires a known normal state, which is not always available. Further improvements in model design would be necessary to reduce the accuracy gap between the standard and pair-wise schemes. Second, while the frequency analysis captured important vowel formants, further refinement is needed to more accurately identify resonant frequencies and their exact positions.  

\section{Ethics}
The study adhered to the guidelines of the
Declaration of Helsinki and was approved by the Taizhou People's Hospital (approval number KY 2023-073-01, obtained on 7 June 2023).

\bibliographystyle{IEEEtran}
\bibliography{mybib}

\begin{thebibliography}{10}
\providecommand{\url}[1]{#1}
\csname url@samestyle\endcsname
\providecommand{\newblock}{\relax}
\providecommand{\bibinfo}[2]{#2}
\providecommand{\BIBentrySTDinterwordspacing}{\spaceskip=0pt\relax}
\providecommand{\BIBentryALTinterwordstretchfactor}{4}
\providecommand{\BIBentryALTinterwordspacing}{\spaceskip=\fontdimen2\font plus
\BIBentryALTinterwordstretchfactor\fontdimen3\font minus \fontdimen4\font\relax}
\providecommand{\BIBforeignlanguage}[2]{{%
\expandafter\ifx\csname l@#1\endcsname\relax
\typeout{** WARNING: IEEEtran.bst: No hyphenation pattern has been}%
\typeout{** loaded for the language `#1'. Using the pattern for}%
\typeout{** the default language instead.}%
\else
\language=\csname l@#1\endcsname
\fi
#2}}
\providecommand{\BIBdecl}{\relax}
\BIBdecl

\bibitem{amir2022remote}
O.~Amir, W.~T. Abraham, Z.~S. Azzam, G.~Berger, S.~D. Anker, S.~P. Pinney, D.~Burkhoff, I.~D. Shallom, C.~Lotan, and E.~R. Edelman, ``Remote speech analysis in the evaluation of hospitalized patients with acute decompensated heart failure,'' \emph{Heart Failure}, vol.~10, no.~1, pp. 41--49, 2022.

\bibitem{murton2023acoustic}
O.~M. Murton, G.~W. Dec, R.~E. Hillman, M.~D. Majmudar, J.~Steiner, J.~V. Guttag, and D.~D. Mehta, ``Acoustic voice and speech biomarkers of treatment status during hospitalization for acute decompensated heart failure,'' \emph{Applied Sciences}, vol.~13, no.~3, p. 1827, 2023.

\bibitem{murton2017acoustic}
O.~M. Murton, R.~E. Hillman, D.~D. Mehta, M.~Semigran, M.~Daher, T.~Cunningham, K.~Verkouw, S.~Tabtabai, J.~Steiner, G.~W. Dec \emph{et~al.}, ``Acoustic speech analysis of patients with decompensated heart failure: a pilot study,'' \emph{The Journal of the Acoustical Society of America}, vol. 142, no.~4, pp. EL401--EL407, 2017.

\bibitem{firmino2023heart}
J.~V. Firmino, M.~Melo, V.~Salemi, K.~Bringel, D.~Leone, R.~Pereira, and M.~Rodrigues, ``Heart failure recognition using human voice analysis and artificial intelligence,'' \emph{Evolutionary Intelligence}, pp. 1--13, 2023.

\bibitem{reddy2021automatic}
M.~K. Reddy, P.~Helkkula, Y.~M. Keerthana, K.~Kaitue, M.~Minkkinen, H.~Tolppanen, T.~Nieminen, and P.~Alku, ``The automatic detection of heart failure using speech signals,'' \emph{Computer Speech \& Language}, vol.~69, p. 101205, 2021.

\bibitem{priyasad2022detecting}
D.~Priyasad, A.~Partovi, S.~Sridharan, M.~Kashefpoor, T.~Fernando, S.~Denman, C.~Fookes, J.~Tang, and D.~Kaye, ``Detecting heart failure through voice analysis using self-supervised mode-based memory fusion,'' in \emph{Proceedings of the 23rd INTERSPEECH Conference}.\hskip 1em plus 0.5em minus 0.4em\relax International Speech Communication Association, 2022, pp. 2848--2852.

\bibitem{mittapalle2022glottal}
K.~R. Mittapalle, H.~Pohjalainen, P.~Helkkula, K.~Kaitue, M.~Minkkinen, H.~Tolppanen, T.~Nieminen, and P.~Alku, ``Glottal flow characteristics in vowels produced by speakers with heart failure,'' \emph{Speech Communication}, vol. 137, pp. 35--43, 2022.

\bibitem{liu2009formant}
H.~Liu and M.~L. Ng, ``Formant characteristics of vowels produced by mandarin esophageal speakers,'' \emph{Journal of voice}, vol.~23, no.~2, pp. 255--260, 2009.

\bibitem{li2018parkinson}
L.~Li, ``Research and implementation of parkinson disease recognition system based on speech recognition,'' Master's thesis, Chongqing University, 2018.

\bibitem{giannitsi20196}
S.~Giannitsi, M.~Bougiakli, A.~Bechlioulis, A.~Kotsia, L.~K. Michalis, and K.~K. Naka, ``6-minute walking test: a useful tool in the management of heart failure patients,'' \emph{Therapeutic advances in cardiovascular disease}, vol.~13, 2019.

\bibitem{eyben2010opensmile}
F.~Eyben, M.~W{\"o}llmer, and B.~Schuller, ``Opensmile: the munich versatile and fast open-source audio feature extractor,'' in \emph{Proceedings of the 18th ACM international conference on Multimedia}, 2010, pp. 1459--1462.

\bibitem{eyben2015geneva}
F.~Eyben, K.~R. Scherer, B.~W. Schuller, J.~Sundberg, E.~Andr{\'e}, C.~Busso, L.~Y. Devillers, J.~Epps, P.~Laukka, S.~S. Narayanan \emph{et~al.}, ``The geneva minimalistic acoustic parameter set (gemaps) for voice research and affective computing,'' \emph{IEEE transactions on affective computing}, vol.~7, no.~2, pp. 190--202, 2015.

\bibitem{schuller2016interspeech}
B.~Schuller, S.~Steidl, A.~Batliner, J.~Hirschberg, J.~K. Burgoon, A.~Baird, A.~Elkins, Y.~Zhang, E.~Coutinho, and K.~Evanini, ``The interspeech 2016 computational paralinguistics challenge: Deception, sincerity \& native language,'' in \emph{17TH Annual Conference of the International Speech Communication Association (Interspeech 2016), Vols 1-5}, vol.~8.\hskip 1em plus 0.5em minus 0.4em\relax ISCA, 2016, pp. 2001--2005.

\bibitem{yang2023attention}
W.~Yang, J.~Liu, P.~Cao, R.~Zhu, Y.~Wang, J.~K. Liu, F.~Wang, and X.~Zhang, ``Attention guided learnable time-domain filterbanks for speech depression detection,'' \emph{Neural Networks}, vol. 165, pp. 135--149, 2023.

\end{thebibliography}

\end{document}